\documentclass[runningheads]{llncs}
\usepackage{multirow}
\usepackage{graphicx}
\usepackage{colortbl}
\definecolor{silver}{rgb}{0.753,0.753,0.753}
\definecolor{mgray}{rgb}{0.128,0.128,0.128}

\mathchardef\mhyphen="2D
\usepackage{fixltx2e}
\usepackage{tikz}
\usepackage{tcolorbox}
\usepackage{CJK}
\usepackage{hyperref}
\usepackage{adjustbox}

\begin{document}
\title{Injecting the BM25 Score as Text Improves BERT-Based Re-rankers}
\author{Arian Askari\inst{1} \and
Amin Abolghasemi\inst{1} \and
Gabriella Pasi\inst{2} \and
Wessel Kraaij\inst{1} \and
Suzan Verberne\inst{1}
}
\authorrunning{A. Askari, A. Abolghasemi, G. Pasi, W.Kraaij, S. Verberne}
\institute{Leiden Institute of Advanced Computer Science, Leiden University \email{\{a.askari,m.a.abolghasemi,w.kraaij,s.verberne\}@liacs.leidenuniv.nl} \and Department of Informatics, Systems and Communication, University of Milano-Bicocca \email{gabriella.pasi@unimib.it}}
\maketitle
\begin{abstract}
In this paper we propose a novel approach for combining first-stage lexical retrieval models and Transformer-based re-rankers: we inject the relevance score of the lexical model as a token in the middle of the input of the cross-encoder re-ranker.
It was shown in prior work that interpolation between the relevance score of lexical and BERT-based re-rankers may not consistently result in higher effectiveness. Our idea is motivated by the finding that BERT models can capture numeric information.
We compare several representations of the BM25 score and inject them as text in the input of four different cross-encoders. We additionally analyze the effect for different query types, and investigate the effectiveness of our method for capturing exact matching relevance. 
Evaluation on the MSMARCO Passage collection and the TREC DL collections shows that the proposed method significantly improves over all cross-encoder re-rankers as well as the common interpolation methods. We show that the improvement is consistent for all query types. We also find an improvement in exact matching capabilities over both BM25 and the cross-encoders.
Our findings indicate that cross-encoder re-rankers can efficiently be improved without additional
computational burden and extra steps in the pipeline by explicitly adding the output of the first-stage ranker to the model input, and this effect is robust for different models and query types. 
\end{abstract}
\section{Introduction}
The commonly used ranking pipeline consists of a first-stage retriever, e.g. BM25 \cite{robertson1994some}, that efficiently retrieves a set of documents from the full document collection, followed by one or more re-rankers \cite{Yan2019IDSTAT,nogueira2019passage} that improve the initial ranking. Currently, the most effective re-rankers are BERT-based rankers with a cross-encoder architecture, concatenating the query and the candidate document in the input \cite{nogueira2019passage,abolghasemi2022improving,hofstatter2020improving,rau2022role}. In this paper, we refer to these re-rankers as Cross-Encoder\textsubscript{CAT} (CE\textsubscript{CAT}). In the common re-ranking set-up, BM25 \cite{robertson1994some} is widely leveraged \cite{anand2021serverless,kamphuis2020bm25,gao2021complement} for finding the top-$k$ documents to be re-ranked; however, the relevance score produced by BM25 based on exact lexical matching is not explicitly taken into account in the second stage. Besides, although cross-encoder re-rankers substantially improve the retrieval effectiveness compared to BM25 alone~\cite{lin2021pretrained}, Rau et al. \cite{rau2022different} show that BM25 is a more effective \textit{exact lexical matcher} than CE\textsubscript{CAT} rankers; in their exact-matching experiment they only use the words from the passage that also appear in the query as the input of the CE\textsubscript{CAT}. This suggests that CE\textsubscript{CAT} re-rankers can be further improved by a better exact word matching, as the presence of query words in the document is one of the strongest signals for relevance in ranking \cite{salton1983introduction,saracevic26review}. Moreover, obtaining improvement in effectiveness by interpolating the scores (score fusion \cite{wu2009applying}) of BM25 and CE\textsubscript{CAT} is challenging: a linear combination of the two scores has shown to decrease effectiveness on the MSMARCO Passage collection compared to only using the CE\textsubscript{CAT} re-ranker in the second stage retrieval \cite{lin2021pretrained}.
\par
\begin{figure}[!t]
   \begin{minipage}{0.48\textwidth}
     \centering
     \includegraphics[width=1.0\linewidth]{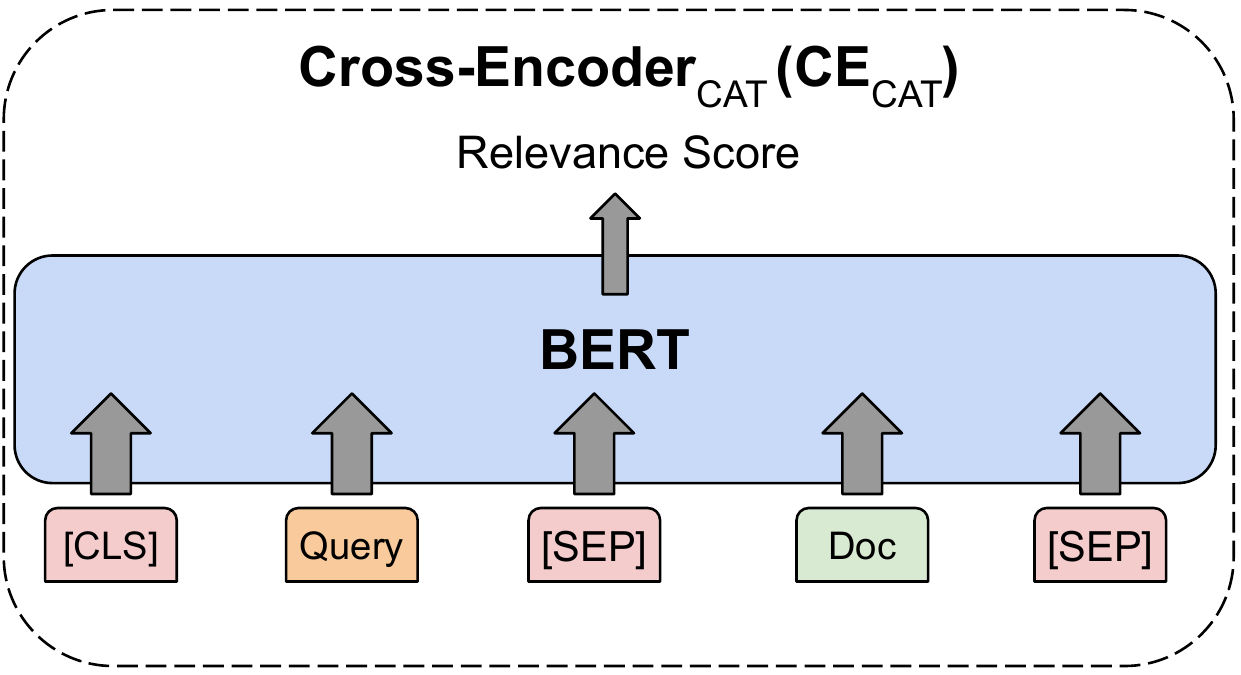}
     \caption{Regular cross-encoder input}\label{fig:bertcat}
   \end{minipage}\hfill
   \begin{minipage}{0.48\textwidth}
     \centering
     \includegraphics[width=1.0\linewidth]{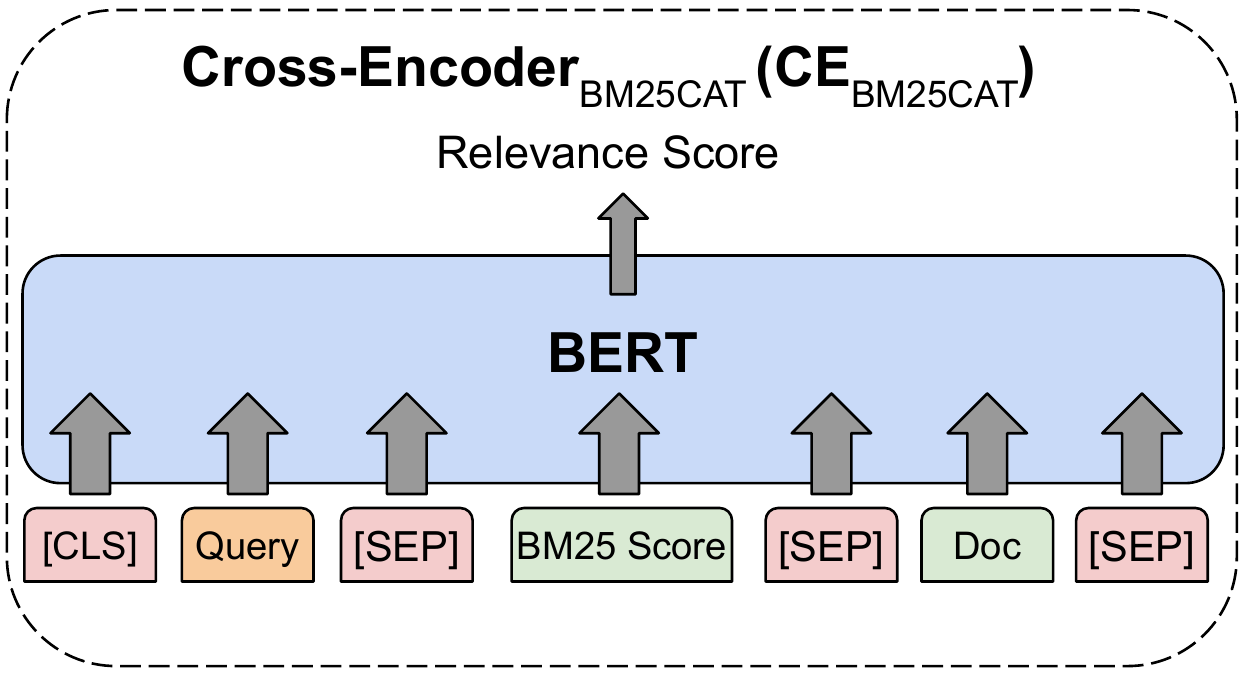}
     \caption{Injection of BM25 in input}\label{fig:bertcatbm25}
   \end{minipage}
\end{figure}

To tackle this problem, in this work, we propose a method to enhance CE\textsubscript{CAT} re-rankers by directly injecting the BM25 score as a string to the input of the Transformer. Figure \ref{fig:bertcatbm25} show our method for the injection of BM25 in the input of the CE re-ranker. We refer to our method as CE\textsubscript{BM25CAT}. 
Our idea is inspired by the finding by Wallace et al. \cite{wallace2019nlp} that BERT models can capture numeracy. In this regard, we address the following research questions:
\par
\textbf{RQ1:} What is the effectiveness of BM25 score injection in addition to the query and document text in the input of CE re-rankers?
\par
\noindent To answer this question we setup two experiments on three datasets: MSMARCO, TREC DL'19 and '20. First, since the BM25 score has no defined range, we investigate the effect of different representations of the BM25 score by applying various normalization methods. We also analyze the effect of converting the normalized scores of BM25 to integers. Second, we evaluate the best representation of BM25 -- based on our empricial study -- on four cross-encoders: BERT-base, BERT-large \cite{vaswani2017attention}, DistillBERT \cite{sanh2019distilbert}, and MiniLM \cite{wang2020minilm}, comparing CE\textsubscript{BM25CAT} to CE\textsubscript{CAT} across different Transformer models with a smaller and larger number of parameters. Next, we compare our proposed approach to common interpolation approaches:
\par
\textbf{RQ2:} What is the effectiveness of CE\textsubscript{BM25CAT} compared to common approaches for combining the final relevance scores of CE\textsubscript{CAT} and BM25?
\par 
\noindent To analyze CE\textsubscript{BM25CAT} and CE\textsubscript{CAT} in terms of exact matching compared to BM25 we address the following question:
\par
\textbf{RQ3:} How effective can CE\textsubscript{BM25CAT} capture exact matching relevance compared to BM25 and CE\textsubscript{CAT}?
\par
\noindent Furthermore, to provide an explanation on the improvement of CE\textsubscript{BM25CAT}, we perform a qualitative analysis of a case where CE\textsubscript{CAT} fails to identify the relevant document that is found using CE\textsubscript{BM25CAT} with the help of the BM25 score.\footnote{In this work, we interchangeably use the words document and passage to refer to unit that should be retrieved.}
\par
To the best of our knowledge, there is no prior work on the effectiveness of cross-encoder re-rankers by injecting a retrieval model's score into their input. Our main contributions in this work are four-fold:
\begin{enumerate}
    \item We provide a strategy for efficiently utilizing BM25 in cross-encoder re-rankers, which yields statistically significant improvements on all official metrics and is verified by thorough experiments and analysis.
    \item We find that our method is more effective than the approaches which linearly interpolate the scores of BM25 and CE\textsubscript{CAT}.
    \item We analyze the exact matching effectiveness of CE\textsubscript{CAT} and CE\textsubscript{BM25CAT} in comparison to BM25. We show that CE\textsubscript{BM25CAT} is a more powerful exact matcher than BM25 while CE\textsubscript{CAT} is less effective than BM25.
    \item We analyze the effectiveness of CE\textsubscript{CAT} and CE\textsubscript{BM25CAT} on different query types. We show that CE\textsubscript{BM25CAT} consistently outperforms CE\textsubscript{CAT} over all type of queries.
\end{enumerate}
\par
\noindent After a discussion of related work in Section~\ref{sec:relwork}, we describe the retrieval models employed in section \ref{sec:retrieval} and the specifics of our experiments and methods in Section \ref{sec:experiments}. The results are examined and the research questions are addressed in Section \ref{sec:results}. Finally, the conclusion is described in Section \ref{sec:conclusion}.
\section{Related work}\label{sec:relwork}
\textbf{Modifying the input of re-rankers. } Boualili et al. \cite{boualili2020markedbert,boualili2022highlighting} propose a method for highlighting exact matching signals by marking the start and the end of each occurrence of the query terms by adding markers to the input. In addition, they modify original passages and expand each passage with a set of generated queries using Doc2query \cite{nogueira2019document} to overcome the vocabulary mismatch problem. This strategy is different from ours in two  aspects: (1) the type of information added to the input: they add four tokens as markers for each occurrence of query terms, adding a burden to the limited input length of 512 tokens for query and document together, while we only add the BM25 score. (2) The need for data augmentation: they need to train a Doc2query model to provide the exact matching signal for improving the BERT re-ranker while our strategy does not need any extra overhead in terms of data augmentation. A few recent, but less related examples are Al-Hajj et al. \cite{al2022arabglossbert}, who experiment with the use of different supervised signals into the input of the cross-encoder to emphasize target words in context and Li et al. \cite{li2022markbert}, who insert boundary markers into the input between contiguous words for Chinese named entity recognition.
\par
\textbf{Numerical information in Transformer models.}
Thawani et al. \cite{thawani2021representing} provide an extensive overview of numeracy in NLP models up to 2021. Wallace et al. \cite{wallace2019nlp} analyze the ability of BERT models to work with numbers and come to the conclusion that the models capture numeracy and are able to do numerical reasoning; however the models appeared to struggle with interpreting floats. Moreover, Zhang et al. \cite{zhang2020language} show that BERT models capture a significant amount of information about numerical scale except for general common-sense reasoning. There are various studies that are inspired by the fact that Transformer models can correctly process numbers \cite{berg2020empirical,muffo-etal-2022-evaluating,johnson2020probing,chen2020numclaim,ConceptNet2022,geva2020injecting}. Gu et al. \cite{gu-budhkar-2021-package} incorporate text, categorical and numerical data as different modalities with Transformers using a combining module accross different classification tasks. They discover that adding tabular features increases the effectiveness while using only text is insufficient and results in the worst performance.
\par
\textbf{Methods for combining rankers.}
Linearly interpolating different rankers' scores has been studied extensively in the literature \cite{lin2021pretrained,wu2009applying,bartell1994automatic,AskariECIR22,askari2021combining}.
In this paper, we investigate multiple linear and non-linear interpolation ensemble methods to analyze the performance of them for combining BM25 and CE\textsubscript{CAT} scores in comparison to CE\textsubscript{BM25CAT}. For the sake of a fair analysis, we do not compare CE\textsubscript{BM25CAT} with a Learning-to-rank approach that is trained on $87$ features by \cite{zhang2021learning}. The use of ensemble methods brings additional overhead in terms of efficiency because it adds one more extra step to the re-ranking pipeline. It is noteworthy to mention that in this paper, we concentrate on analyzing the improvement by combining the first-stage retriever and a BERT-based re-ranker: BM25 and CE\textsubscript{CAT} respectively. However, we are aware that combining scores of BM25 and Dense Retrievers that both are first-stage retrievers has also shown improvements \cite{ZucconInterpolation,abolghasemiICTIR,althammer2021dossier} that are outside the scope of our study. In particular, CLEAR \cite{gao2021complement} proposes an approach to train the dense retrievers to encode semantics that BM25 fails to capture for first stage retrieval. However, in this study, our aim is to improve re-ranking in the second stage of two-stage retrieval setting.
\section{Methods}\label{sec:retrieval}
\subsection{First stage ranker: BM25}
Lexical retrievers estimate the relevance of a document to a query based on word overlap \cite{robertson2009probabilistic}.
Many lexical methods, including vector space models, Okapi BM25, and query likelihood, have been developed in previous decades. We use BM25 because of its popularity as first-stage ranker in current systems. Based on the statistics of the words that overlap between the query and the document, BM25 calculates a score for the pair:
\begin{equation}
s_{lex}(q,d) = BM25(q,d) = \sum_{t \in q \cap d }{rsj_t . \frac{tf_{t,d}}{tf_{t,d} + k_{1} \{ (1-b) + b \frac{|d|}{l} \} }}
\end{equation}
where $t$ is a term, $tf_{t,d}$ is the frequency of $t$ in document $d$, $rsj_t$ is the Robertson-Spärck Jones weight \cite{robertson1994some} of $t$, and $l$ is the average document length. $k_1$ and $b$ are parameters \cite{pyseriniTunedBm25,Lin_etal_SIGIR2021_Pyserini}.
\subsection{CE\textsubscript{CAT}: cross-encoder re-rankers without BM25 injection}
Concatenating query and passage input sequences is the typical method for using cross-encoder (e.g., BERT) architectures with pre-trained Transformer models in a re-ranking setup \cite{nogueira2019passage,macavaney2019cedr,yilmaz2019cross,hofstatter2020improving}. This basic design is referred to as CE\textsubscript{CAT} and shown in Figure \ref{fig:bertcat}. The query $q_{1:m}$ and passage $p_{1:n}$ sequences are concatenated with the $[SEP]$ token, and the $[CLS]$ token representation computed by CE is scored with a single linear layer $W_s$ in the CE\textsubscript{CAT} ranking model:
\begin{equation}
    CE_{CAT}(q_{1:m},p_{1:n}) = CE([CLS]\,q\,[SEP]\,p\,[SEP]) * W_s
\end{equation}
We use CE\textsubscript{CAT} as our baseline re-ranker architecture. We evaluate different cross-encoder models in our experiments and all of them follow the above design.
\subsection{CE\textsubscript{BM25CAT}: cross-encoder re-rankers with BM25 injection}\label{subsec:CEBM25CAT}
To study the effectiveness of injecting the BM25 score into the input, we modify the input of the basic input format as follows and call it CE\textsubscript{BM25CAT}:
\begin{equation}
    CE_{BM25CAT}(q_{1:m},p_{1:n})=CE([CLS]\,q\,[SEP]\,BM25\,[SEP]\,p\,[SEP]) * W_s
\end{equation}
where BM25 represent the relevance score produced by BM25 between query and passage.
\par
We study different representations of BM25 to find the optimal approach for injecting BM25 into the cross-encoders. The reasons are: $(1)$ BM25 scores do not have an upper bound and should be normalized for having an interpretable score given a query and passage; $(2)$ BERT-based models can process integers better than floating point numbers \cite{wallace2019nlp} so we analyze if converting the normalized score to an integer is more effective than injecting the floating point score. For normalizing BM25 scores, we compare three different normalization methods: Min-Max, Standardization (Z-score), and Sum:
\begin{equation}
    Min \mhyphen Max(s_{BM25}) = \frac{s_{BM25}-s_{min}}{s_{max}-s_{min}} 
\end{equation}
\begin{equation}
    Standard(s_{BM25}) = \frac{s_{BM25}- \mu(S)}{\sigma(S)} 
\end{equation}
\begin{equation}
    Sum(s_{BM25}) = \frac{s_{BM25}}{sum(S)} 
\end{equation}
Where $s_{BM25}$ is the original score, and $s_{max}$ and $s_{min}$ are the maximum and minimum scores respectively, in the ranked list. $Sum(S)$, $\mu(S)$, and $\sigma(S)$ refer to sum, average and standard deviation over the scores of all passages retrieved for a query. The anticipated effect of the Sum normalizer is that the sum of the scores of all passages in the ranked list will be $1$; thus, if the top-$n$ passages receive much higher scores than the rest, their normalized scores will have a larger difference with the rest of passages' scores in the ranked list; this distance could give a good signal to CE\textsubscript{BM25CAT}.
We experiment with Min-Max and Standardization in a local and a global setting. In the local setting, we get the minimum or maximum (for Min-Max) and mean and standard deviation (for Standard) from the ranked list of scores per query. In the global setting, we use $\{0,50,42,6\}$ as \{minimum, maximum, mean, standard deviation\} as they have been empirically suggested in prior work to be used as default values across different queries to globally normalize BM25 scores \cite{ltrsolr}. In our data, the \{minimum, maximum, mean, standard deviation\} values are $\{0,98,7,5\}$ across all queries. Because of the differences between the recommended defaults and the statistics of our collections, we explore other global values for Min-Max, using $25,50,75,100$ as maximum and $0$ as minimum. However, we got the best result using default values of \cite{ltrsolr}.
To convert the float numbers to integers we multiply the normalized score to $100$ and discard decimals. Finally, we store the number as a string.
\subsection{Linear interpolation ensembles of BM25 and CE\textsubscript{CAT}}\label{sec:experiments_ensemble}
We compare our approach to common ensemble methods \cite{lin2021pretrained,combiningScoresJimmyLin2021} for interpolating BM25 and BERT re-rankers. We combine the scores linearly using the following methods:
$(1)$ Sum: compute sum over BM25 and CE\textsubscript{CAT} scores, 
$(2)$ Max: select maximum between BM25 and CE\textsubscript{CAT} scores, and 
$(3)$ Weighted-Sum: 
\begin{equation}
    s_i = \alpha \, . \, s_{BM25} + (1-\alpha) \, . \, s_{CE_{CAT}}
\end{equation}
Where $s_i$ is the weighted sum produced by the interpolation, $s_{BM25}$ is the normalized BM25 score, $s_{CE_{CAT}}$ is the CE\textsubscript{CAT} score, and $\alpha \in [0..1]$ is a weight that indicates the relative importance. Since CE\textsubscript{CAT} score $\in [0,1]$, we also normalize BM25 score using Min-Max normalization. Furthermore, we train ensemble models that take $s_{BM25}$ and $s_{CE_{CAT}}$ as features. We experiment with three different classifiers for this purpose: SVM with a linear kernel, SVM with an RBFkernel, Naive Bayes, and Multi Layer Perceptron (MLP) as a non-linear method and report the best classifier performance in Section \ref{sec:result_ensemble}.
\section{Experimental design}\label{sec:experiments}
\textbf{Dataset and metrics.}
We conduct our experiments on the MSMARCO-passage collection \cite{nguyen2016ms} and the two TREC Deep Learning tracks (TREC-DL'19 and TREC-DL'20) \cite{craswell2020overview,craswell2021overview}. The MSMARCO-passage dataset contains about $8.8$ million passages (average length: $73.1$ words) and about $1$ million natural language queries (average length: $7.5$ words) and has been extensively used to train deep language models for ranking because of the large number of queries. Following prior work on MSMARCO \cite{khattab2020colbert,lin2021pretrained,macavaney2020expansion,zhuang2021tilde,zhuang2021deep}, we use the dev set ($\sim 7k$ queries) for our empirical evaluation. $MAP@1000$ and $nDCG@10$ are calculated in addition to the official evaluation metric $MRR@10$. The passage corpus of MSMARCO is shared with TREC DL'19 and DL'20 collections with $43$ and $54$ queries respectively. We evaluate our experiments on these collections using $nDCG@10$ and $MAP@1000$, as is standard practice in TREC DL \cite{craswell2021overview,craswell2020overview} to make our results comparable to previously published and upcoming research. We cap the query length at $30$ tokens and the passage length at $200$ tokens following prior work \cite{hofstatter2020improving}. 
\par
\textbf{Training configuration and model parameters.}
We use the Huggingface library \cite{wolf2019huggingface}, Cross-encoder package of Sentence-transformers library \cite{reimers-2019-sentence-bert}, and PyTorch \cite{paszke2017automatic} for the cross-encoder re-ranking training and inference.
For injecting the BM25 score as text, we pass the BM25 score in string format into the BERT tokenizer in a similar way to passing query and document. Please note that the integer numbers are already included in the BERT tokenizer's vocabulary, allowing for appropriate tokenization.
Following prior work \cite{hofstatter2020improving} we use the Adam \cite{kingma2014adam} optimizer with a learning rate of $7*10^{-6}$ for all cross-encoder layers, regardless of the number of layers trained. To train cross-encoder re-rankers for each TREC DL collection, we use the other TREC DL query set as the validation set and we select both TREC DL ('19 and '20) query sets as the validation set to train CEs for the MSMARCO Passage collection. We employ early stopping, based on the nDCG@10 value of the validation set. We use a training batch size of 32. For all cross-Encoder re-rankers, we use Cross-Entropy loss \cite{zhang2018generalized} . For the lexical retrieval with BM25 we employ the tuned parameters from the Anserini documentation \cite{pyseriniTunedBm25,Lin_etal_SIGIR2021_Pyserini}. \footnote{The code is available on \href{https://github.com/arian-askari/injecting\_bm25\_score\_bert\_reranker}{https://github.com/arian-askari/injecting\_bm25\_score\_bert}}
\section{Results}\label{sec:results}
\subsection{Main results: addressing our research questions}
\paragraph{\textbf{Choice of BM25 score representation}}\label{subsec:inputformat}
As introduced in Section \ref{subsec:CEBM25CAT}, we compare different representations of the BM25 score in Table \ref{tab:represent_bm25} for injection into CE\textsubscript{BM25CAT}. We chose MiniLM \cite{wang2020minilm} for this study as it has shown competitive results in comparison to BERT-based models while it is $3$ times smaller and $6$ times faster.
\footnote{\url{https://huggingface.co/microsoft/MiniLM-L12-H384-uncased}}
Our first interesting observation is that injecting the original float score rounded down to 2 decimal points (row $b$) of BM25 into the input seems to slightly improve the effectiveness of re-ranker. We assume this is due to the fact that the average query and passage length is relatively small in the MSMARCO Passage collection, which prevents from getting high numbers -- with low interpretability for BERT -- as BM25 score. Second, we find that the normalized BM25 score with Min-Max in the global normalization setting converted to integer (row $f$) is the most significant effective\footnote{Although the evaluation metrics are not in an interval scale, Craswell et al. \cite{craswell2021ms} show that they are mostly reliable in practice on MSMARCO for statistical testing.} representation for injecting BM25.
\begin{table}[ht]
    \centering
    \caption{Effectiveness results. Lines $b$-$n$ refer to the MiniLM\textsubscript{BMCAT} re-ranker using different representations of the BM25 score as text. Significance is shown with $\dagger$ for the best result (row $f$) compared to MiniLM\textsubscript{CAT} (row $a$). Statistical significance was measured with a paired t-test ($p<0.05$) with Bonferroni correction for multiple testing.}
    \arrayrulecolor[rgb]{0.753,0.753,0.753}
    \resizebox{12cm}{!}{
         \begin{tabular}{ll|c|c|ccc} 
             \arrayrulecolor{black}\hline
            \rule{0pt}{4ex}&\textbf{Normalization\,\,} & \multicolumn{1}{l!{\color{silver}\vrule}}{\textbf{\,Local/Global\,}} & \textbf{\,Float/Integer\,} & \multicolumn{3}{c}{\textbf{MSMARCO DEV}} \\
             \rule{0pt}{3ex}\rule[-2ex]{0pt}{0pt}& & \multicolumn{1}{l!{\color{silver}\vrule}}{} & & \, nDCG@10 & \, MAP \, & \, MRR@10 \\ 
             \hline
             \rule[-2ex]{0pt}{0pt}\rule{0pt}{3ex}$(a)$& \multicolumn{3}{l!{\color{silver}\vrule}}{MiniLM\textsubscript{CAT} $($without injecting BM25 score$)$} & \multicolumn{1}{c}{.419} & \multicolumn{1}{c}{.363} & \multicolumn{1}{c}{.360} \\ \arrayrulecolor{black}\hline
             \rule[-2ex]{0pt}{0pt}\rule{0pt}{3ex}$(b)$& Original Score & —-- & —-- & .420 & .364 & .362 \\ 
             \arrayrulecolor[rgb]{0.753,0.753,0.753}\hline
             \rule{0pt}{3ex}$(c)$& Min-Max & Local & Float & .411 & .359 & .354 \\
             $(d)$ & Min-Max & Local & Integer & .414 & .361 & .355 \\
             $(e)$ & Min-Max & Global & Float & .422 & .365 & .363  \\
             \rule[-2ex]{0pt}{0pt}$(f)$ & Min-Max & Global & Integer & \textbf{.424}$\dagger$ & \textbf{.368}$\dagger$ & \textbf{.367}$\dagger$ \\ 
             \hline
             \rule{0pt}{3ex}$(g)$ & Standard & Local & Float & .407 & .355 & .352 \\
             $(h)$ & Standard & Local & Integer & .410 & .358 & .354 \\
             $(i)$ & Standard & Global & Float & .420 & .363 & .361  \\
             \rule[-2ex]{0pt}{0pt}$(j)$ & Standard & Global & Integer & .421 & .365 & .363 \\ 
             \hline
             \rule{0pt}{3ex}$(k)$ & Sum & - & Float &  .402 & .349 & .338 \\
             \rule[-2ex]{0pt}{0pt}$(l)$ & Sum & - & Integer & .405 & .350 & .342 \\
             \arrayrulecolor{black}\hline
         \end{tabular}
         \label{tab:represent_bm25}
     }
\end{table}

The global normalization setting gives better results for both Min-Max (rows $e,f$) and Standardization (rows $i,j$) than local normalization (rows $c,d$ and $g,h$).\footnote{The range of normalized integer scores using the best normalizer (row $f$) are from $0$ to $196$ as the maximum BM25 score in the collection is $98$.}
The reason is probably that in the global setting a candidate document obtains a high normalized score (close to $1$ in the floating point representation) if its original score is close to default maximum (for Min-Max normalization) so the normalized score could be more interpretable across different queries. On the other hand, in the local setting, the passages ranked at position 1 always receive $1$ as normalized score with Min-Max even if its original score is not high and it does not have a big difference with the last passage in the ranked list.
\par
\begin{figure}[t]
\centering
\scalebox{0.99}{
  \includegraphics[width=\linewidth]{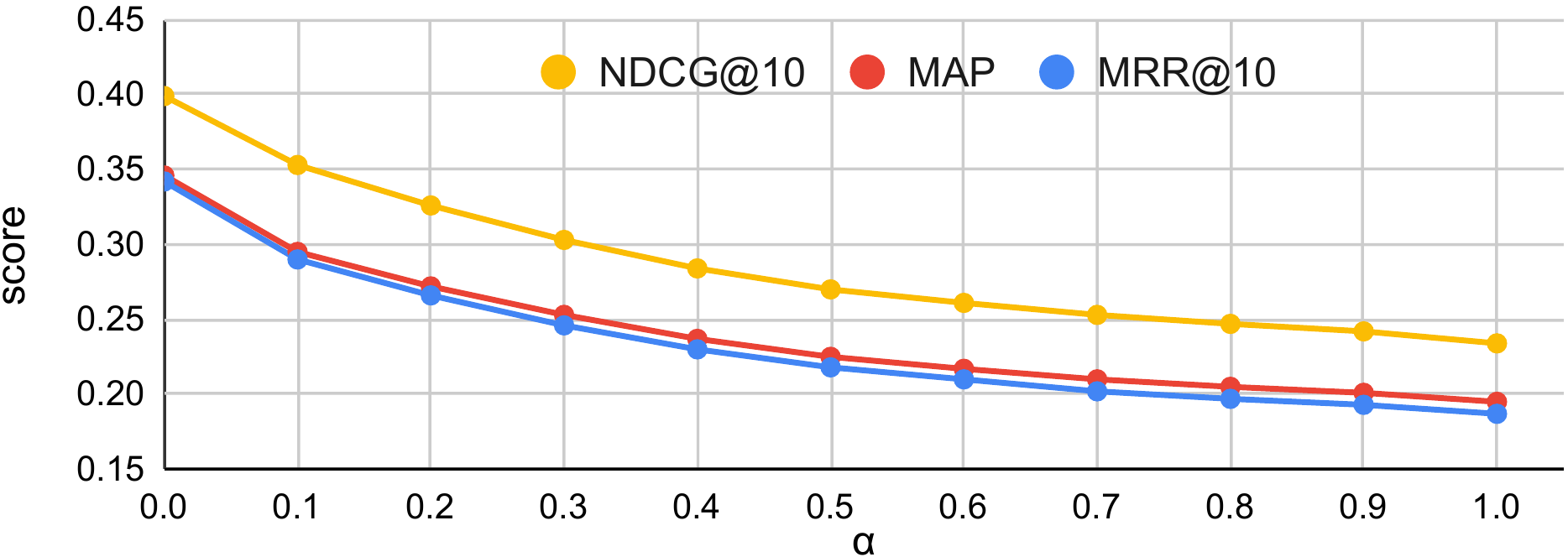}
}
  \caption{Effectiveness on MSMARCO DEV with varying the interpolation weight of BM25 and BERT-Base\textsubscript{CAT} scores. $\alpha=0$ means only BERT\textsubscript{CAT} scores are used. }
  \label{fig:ensemble}
\end{figure}

Moreover, converting the normalized float score to integers gives better results for both Min-Max (rows $d,f$) and Standardization (rows $h,j$) than the float representation (rows $c,e$ and $g,i$). We find that Min-Max normalization is a better representation for injecting BM25 than Standardization, which could be due to the fact that in Min-Max the normalized score could not be negative, and, as a result, interpreting  the injected score is easier for CE\textsubscript{BM25CAT}. We find that the Sum normalizer (rows $k$ and $l$) decreases effectiveness. Apparently, our expectation that Sum would help distinguish between the top-$n$ passages and the remaining passages in the ranked list (see Section~\ref{subsec:inputformat}) is not true.
\begin{table}[t]
\centering
\caption{Effectiveness results. Fine-tuned cross-encoders are used for re-ranking over BM25 first stage retrieval with a re-ranking depth of $1000$. $\dagger$ indicates a statistically significant improvement of a cross-encoder with BM25 score injection as text into the input (Cross-encoder\textsubscript{BM25CAT}) over the same cross-encoder without BM25 score injection (Cross-encoder\textsubscript{CAT}). Statistical significance was measured with a paired t-test ($p<0.05$) with Bonferroni correction for multiple testing.
}
\arrayrulecolor[rgb]{0.753,0.753,0.753}
\resizebox{12cm}{!}{
    \begin{tabular}{ll|cc|cc|ccc} 
    \arrayrulecolor{black}\hline
     & \rule{0pt}{4ex}\textbf{Model} & \multicolumn{2}{c!{\color{silver}\vrule}}{\textbf{TREC DL 20}} & \multicolumn{2}{c!{\color{silver}\vrule}}{\textbf{TREC DL 19}} & \multicolumn{3}{c}{\textbf{MSMARCO DEV}} \\
     & \rule[-3ex]{0pt}{0pt} \rule{0pt}{3ex} & \, nDCG@10   &    \, MAP \, & \, nDCG@10 & \, MAP \, & \, nDCG@10 & \, MAP \, & \, MRR@10 \\ 
    \hline
     & \rule[-3ex]{0pt}{0pt}\rule{0pt}{4ex}BM25 & .480 & .286 & .506 & .377 & .234 & .195 & .187 \\  
    \hline
    \multicolumn{2}{l!{\color{silver}\vrule}}{\rule{0pt}{3ex}\textbf{Re-rankers}} & & & & & & & \\
     & \rule{0pt}{4ex}BERT-Base\textsubscript{CAT} & .689 & .447 & .713 & .441 & .399 & .346 & .342 \\
     & \rule[-3ex]{0pt}{0pt}BERT-Base\textsubscript{BM25CAT} & .705$\dagger$ & .475$\dagger$ & .723$\dagger$ & .453$\dagger$ & .422$\dagger$ & .367$\dagger$ & .364$\dagger$ \\ 
    \arrayrulecolor[rgb]{0.502,0.502,0.502}\hline
     & \rule{0pt}{4ex}BERT-Large\textsubscript{CAT} & .695 & .464 & .714 & .467 & .401 & .344 & .360 \\
     & \rule[-3ex]{0pt}{0pt}BERT-Large\textsubscript{BM25CAT} & \textbf{.728}$\dagger$ & \textbf{.482}$\dagger$ & \textbf{.731}$\dagger$ & \textbf{.477}$\dagger$ & \textbf{.424}$\dagger$ & .367$\dagger$ & \textbf{.369}$\dagger$ \\ 
    \hline
     & \rule{0pt}{4ex}DistilBERT\textsubscript{CAT} & .670 & .442 & .679 & .440 & .383 & .310 & .325\\
     & \rule[-3ex]{0pt}{0pt}DistilBERT\textsubscript{BM25CAT} & .682$\dagger$ & .456$\dagger$ & .699$\dagger$ & .451$\dagger$ & .390$\dagger$ & .323$\dagger$ & .339$\dagger$ \\ 
    \hline
     & \rule{0pt}{4ex}MiniLM\textsubscript{CAT} & .681 & .448 & .704 & .452 & .419 & .363 & .360 \\
     & \rule[-3ex]{0pt}{0pt}MiniLM\textsubscript{BM25CAT} & .710$\dagger$ & .473$\dagger$ & .711$\dagger$ & .463$\dagger$ & \textbf{.424}$\dagger$ & \textbf{.368}$\dagger$ &  .367$\dagger$ \\
     \arrayrulecolor{black}\hline
    \end{tabular}
    \label{tab:effectiveness}
}
\arrayrulecolor{black}
\end{table}
\paragraph{\textbf{Impact of BM25 injection for various cross-encoders (RQ1)}}
Table~\ref{tab:effectiveness} shows that injecting the BM25 score -- using the best normalizer which is  Min-Max in the global normalization setting converted to integer -- into all four cross-encoders improves their effectiveness in all of the metrics compared to using them without injecting BM25. This shows that injecting the BM25 score into the input as a small modification to the current re-ranking pipeline improves the re-ranking effectiveness. This is without any additional computational burden as we train CE\textsubscript{CAT} and CE\textsubscript{BM25CAT} in a completely equal setting in terms of number of epochs, batch size, etc. We receive the highest result by BERT-Large\textsubscript{BM25CAT} for cross-encoder with BM25 injection, which could be due to the higher number of parameters of the model. We find that the results of MiniLM are similar to those for BERT-Base on MSMARCO-DEV while the former is more efficient. 
\begin{table}[t]
\centering
\caption{The effectiveness of injecting BM25 score into the input (Bert-Base\textsubscript{BM25CAT}) compared to interpolation performance of BM25 and Bert-Base\textsubscript{CAT} using common ensemble methods.
}
\arrayrulecolor[rgb]{0.753,0.753,0.753}
    \resizebox{12cm}{!}{
        \begin{tabular}{ll|l|ccc} 
        \arrayrulecolor{black}\hline
         & \multirow{2}{*}{\, Model\, } & \multirow{2}{*}{\, Ensemble} & \multicolumn{3}{c}{\textbf{\rule[-2ex]{0pt}{0pt}\rule{0pt}{3ex}MSMARCO DEV}} \\
         & & & \rule[-2ex]{0pt}{0pt} \, nDCG@10 & \, MAP \, & \, MRR@10 \\ 
        \arrayrulecolor{silver}\hline
         \rule{0pt}{3ex}\, & \, BM25 & \multicolumn{1}{l!{\color{silver}\vrule}}{\, ---} & .234  & .195 & .187 \\
         \rule[-2ex]{0pt}{0pt}\, & \, BERT-Base\textsubscript{CAT}  & \multicolumn{1}{l!{\color{silver}\vrule}}{\, ---} & .399  & .346 & .342 \\
         \arrayrulecolor{silver}\hline
         \rule{0pt}{3ex}\, & \, BM25 and BERT-Base\textsubscript{CAT}  & \multicolumn{1}{l!{\color{silver}\vrule}}{\, Sum} & .270  & .225 & .218  \\
         \, & \, BM25 and BERT-Base\textsubscript{CAT}  & \multicolumn{1}{l!{\color{silver}\vrule}}{\, Max} & .237  & .197 & .190  \\
         \rule[-2ex]{0pt}{0pt}\, & \, BM25 and BERT-Base\textsubscript{CAT}  & \multicolumn{1}{l!{\color{silver}\vrule}}{\, Weighted-Sum (tuned)}  & .353  & .295 & .290   \\
        \arrayrulecolor{mgray}\hline
        \rule[-2ex]{0pt}{0pt}\rule{0pt}{3ex}\, & \, BM25 and BERT-Base\textsubscript{CAT}  & \multicolumn{1}{l!{\color{silver}\vrule}}{\, Naive Bayes} & .314 & .260 & .254  \\
        \arrayrulecolor{mgray}\hline
        
         \rule[-2ex]{0pt}{0pt}\rule{0pt}{3ex}& \, BERT-Base\textsubscript{BM25CAT} & \multicolumn{1}{l!{\color{silver}\vrule}}{\, BM25 Score Injection} & \textbf{.422}  & \textbf{.367} & \textbf{.364}  \\
        \arrayrulecolor{black}\hline
        \end{tabular}
    }
\label{table:ensemble}
\end{table}
\paragraph{\textbf{Comparing BM25 Injection with Ensemble Methods (RQ2)}} \label{sec:result_ensemble}
Table \ref{table:ensemble} shows that while injecting BM25 leads to improvement, regular ensemble methods and Naive Bayes classifier fail to do so; combining the scores of BM25 and BERT\textsubscript{CAT} in a linear and non-linear (MLP) interpolation ensemble setting even leads to lower effectiveness than using the cross-encoder as sole re-ranker. Therefore, our strategy is a better solution than linear interpolation. We only report results for Naive Bayes -- having BM25 and BERT\textsubscript{CAT} score as features -- as it had the highest effectiveness of the four estimators. Still, the effectiveness is much lower than  BERT\textsubscript{BM25CAT} and also lower than a simple Weighted-Sum. Weighted-Sum (tuned) in Table \ref{table:ensemble} is tuned on the validation set, for which  $\alpha=0.1$ was found to be optimal. We analyze the effect of different $\alpha$ values in a weighted linear interpolation (Weighted-Sum) to draw a more complete picture on the impact of combining scores on the DEV set. Figure \ref{fig:ensemble} shows that by increasing the weight of BM25, the effectiveness decreases. The figure also shows that the tuned alpha which was found on the validation set in Table \ref{table:ensemble} is not the most optimal possible alpha value for the DEV set. The highest effectiveness for $\alpha=0.0$ in Figure \ref{fig:ensemble} confirms we should not combine the scores by current interpolation methods and only using scores of Bert-Base\textsubscript{CAT} is better, at least for the MSMARCO passage collection.
\paragraph{\textbf{Exact Matching Relevance Results (RQ3)}}\label{sec:result_exact}
To conduct exact matching analysis, we replace the passage words that do not appear in the query with the $[MASK]$ token, leaving the model only with a skeleton of the original passage and force it to rely on the exact word matches between query and passage \cite{rau2022different}.
We do not train models on this input but use our models that were fine-tuned on the original data.
Table \ref{tab:exact_matching} shows that BERT-Base\textsubscript{BM25CAT} performs better than both BM25 and BERT-Base\textsubscript{CAT} in the exact matching setting on all metrics.
Moreover, we found that the percentage of relevant passages ranked in top-$10$ that are common between BM25 and BERT\textsubscript{BM25CAT} is $40\%$, which is higher than the percentage of relevant passages between BM25 and BERT\textsubscript{CAT} ($37\%$). Therefore, the higher effectiveness of BERT\textsubscript{BM25CAT} in exact matching setting could be at least partly because it mimics BM25 more than BERT\textsubscript{CAT}. In comparison, this percentage is $57$ between BERT\textsubscript{BM25CAT} and BERT\textsubscript{CAT}.
\begin{table}[]
\centering
\caption{Comparing exact matching effectiveness of BERT-Base\textsubscript{BM25CAT} and BERT-Base\textsubscript{CAT} by keeping only the query words in each passage for re-ranking. The increase and decrease of effectiveness compared to BM25 is indicated with $\uparrow$ and $\downarrow$.}
\arrayrulecolor[rgb]{0.753,0.753,0.753}
    \resizebox{12cm}{!}{
        \begin{tabular}{ll|l|ccc} 
        \arrayrulecolor{black}\hline
         & \multirow{2}{*}{\, Model} & \multirow{2}{*}{\, Input} & \multicolumn{3}{c}{\textbf{\rule[-2ex]{0pt}{0pt}\rule{0pt}{3ex}MSMARCO DEV}} \\
         & & & \rule[-2ex]{0pt}{0pt} \, nDCG@10 & \, MAP \, & \, MRR@10 \\ 
        \arrayrulecolor[rgb]{0.502,0.502,0.502}\hline
        \rule{0pt}{3ex}\, & \, BM25 & \multicolumn{1}{l!{\color{silver}\vrule}}{\, Full text} & .234  & .195 & .187 \\
        \, & \, BERT-Base\textsubscript{CAT} & \, Only query words\, &  \,\,.218  ($\downarrow$1.6)  & .186 ($\downarrow$0.9) & .180 ($\downarrow$0.7) \\
        \rule[-2ex]{0pt}{0pt}\, & \, BERT-Base\textsubscript{BM25CAT} & \, Only query words &\textbf{.243} ($\uparrow$.9)\, & \,\textbf{.209}  ($\uparrow$1.4)\, & \textbf{.202} ($\uparrow$1.5) \\ 
        \hline
        \arrayrulecolor{black}\hline
        \end{tabular}
    }
\label{tab:exact_matching}
\end{table}
\subsection{Analysis of the results}
\textbf{Query types.}
In order to analyze the effectiveness of  BERT-base\textsubscript{CAT} and BERT-base\textsubscript{BM25CAT} across different types of questions, we classify questions based on the lexical answer type. We use the rule-based answer type classifier\footnote{\url{https://github.com/superscriptjs/qtypes}} inspired by \cite{li2002learning} to extract answer types. We classify MSMARCO queries into 6 answer types: abbreviation, location, description, human, numerical and entity. $4105$ queries have a valid answer type and at least one relevant passage in the top-$1000$.
We perform our analysis in two different settings: normal (full-text) and exact-matching (keeping only query words and replacing non-query words with $[MASK]$). The average $MRR@10$ per query type is shown in Table \ref{tab:query_type}. The table shows that BERT\textsubscript{BM25CAT} is more effective than BERT\textsubscript{CAT} consistently on all types of queries.
\begin{table}[t]
\centering
\caption{MRR@10 on MSMARCO-DEV per query type for comparing  BERT-Base\textsubscript{BM25CAT} and BERT-Base\textsubscript{CAT} on different query types in full-text and exact-matching (only keeping query words) settings.}
\arrayrulecolor[rgb]{0.753,0.753,0.753}
    \resizebox{12cm}{!}{
        \begin{tabular}{ll|l|cccccc} 
        \arrayrulecolor{black}\hline
         & \rule{0pt}{4ex}Model &\,Input & \multicolumn{1}{c}{\, ABBR\, } & LOC\, & DESC\, & HUM\, & NUM\, & ENTY\, \\ 
        \arrayrulecolor[rgb]{0.502,0.502,0.502}\hline
        & \# \rule{0pt}{4ex}queries       & \multicolumn{1}{l!{\color{silver}\vrule}}{}                 & \multicolumn{1}{c}{9}    & \multicolumn{1}{c}{493}  & \multicolumn{1}{c}{1887} & 455  & 933  & 328   \\
        \arrayrulecolor[rgb]{0.753,0.753,0.753}
        & BERT-BaseCAT & \,Full text\,  & .574 & .477 & .397 & .435 & .361 & .399 \\
         & \rule[-2ex]{0pt}{0pt}BERT-BaseBM25CAT &\,Full text\,  & \textbf{.592} & \textbf{.503} & \textbf{.428} & \textbf{.457} & \textbf{.405} & \textbf{.411} \\ 
        \hline
         &\rule{0pt}{4ex}BM25 & \multicolumn{1}{l!{\color{silver}\vrule}}{\,Only query words\,} & .184 & .256 & .215 & .238 & .200 & \textbf{.221} \\
        & BERT-BaseCAT & \,Only query words\, & .404 & .204 & .224 & .240 & .177 & .200 \\
        & \rule[-2ex]{0pt}{0pt}BERT-BaseBM25CAT & \multicolumn{1}{l!{\color{silver}\vrule}}{\,Only query words\,} & \textbf{.438} & \textbf{.278} & \textbf{.245} & \textbf{.258} & \textbf{.215} & .216 \\
        \arrayrulecolor{black}\hline
        \end{tabular}
    }
\label{tab:query_type}
\end{table}
\begin{figure}
\scriptsize
\centering
\caption{Example query and two passages in the input of BERT\textsubscript{BM25CAT}. The color of each word indicates the word-level attribution value according to Integrated Gradient (IG) \cite{sundararajan2017axiomatic}, where red is positive, blue is negative, and white is neutral. We use the brightness of different colors to indicate the values of these gradients. }
\arrayrulecolor[rgb]{0.753,0.753,0.753}
\scalebox{0.80}{
    \begin{tabular}{p{11cm}|c|p{3cm}} 
    \arrayrulecolor{black}\hline
    \rule{0pt}{3ex}\rule[-2ex]{0pt}{0pt}Query [SEP] BM25 [SEP] Passage & Label & Model: Rank \\ 
    \arrayrulecolor[rgb]{0.502,0.502,0.502}\hline
    \rule{0pt}{4ex}\colorbox{red!0.0}{\strut [CLS]} \colorbox{blue!40.91088089009047}{\strut what} \colorbox{blue!34.98627128288114}{\strut is} \colorbox{blue!28.836949285746034}{\strut the} \colorbox{blue!37.2}{\strut shingles} \colorbox{red!23.7121748}{\strut jab} \colorbox{red!27.446856038961254}{\strut ?} \colorbox{red!0.0}{\strut [SEP]} \colorbox{blue!59.041815520803844}{\strut 22} \colorbox{blue!16.72737963183179}{\strut [SEP]} \colorbox{blue!11.27764064464184}{\strut the} \colorbox{red!39.40945}{\strut shingles} \colorbox{blue!27.43521052787053}{\strut vaccine} \colorbox{blue!14.468601692001586}{\strut .} \colorbox{blue!27.77131472749809}{\strut the} \colorbox{blue!19.19327067598481}{\strut vaccine} \colorbox{red!13.894921543793497}{\strut ,} \colorbox{red!14.9827478652407}{\strut called} \colorbox{blue!2.1519884}{\strut zostavax} \colorbox{red!8.821212931728395}{\strut ,} \colorbox{red!15.94387850700808}{\strut is} \colorbox{blue!2.3739219436855827}{\strut given} \colorbox{red!3.541043392744023}{\strut as} \colorbox{red!8.232288341442084}{\strut a} \colorbox{red!8.227893243683223}{\strut single} \colorbox{blue!12.407899625388746}{\strut injection} \colorbox{red!2.910927401598767}{\strut under} \colorbox{blue!0.19742208742405165}{\strut the} \colorbox{blue!3.371029939964714}{\strut skin} \colorbox{blue!1.720481149118187}{\strut (} \colorbox{red!10.5491942}{\strut subcutaneously} \colorbox{red!1.576029432259837}{\strut )} \colorbox{red!7.17710465553816}{\strut .} \colorbox{red!6.429270805438332}{\strut it} \colorbox{red!7.88615104209289}{\strut can} \colorbox{red!5.04978782881975}{\strut be} \colorbox{red!5.424809781278258}{\strut given} \colorbox{red!10.253032467609534}{\strut at} \colorbox{red!1.3124829352642469}{\strut any} \colorbox{blue!2.821835456427737}{\strut time} \colorbox{blue!0.15540499204660282}{\strut in} \colorbox{blue!4.789388174158487}{\strut the} \colorbox{red!1.3682692735729773}{\strut year} \colorbox{red!24.445497242056174}{\strut .} \colorbox{blue!4.898805720830903}{\strut unlike} \colorbox{red!9.019066827167837}{\strut with} \colorbox{blue!6.9012954095608965}{\strut the} \colorbox{blue!11.995395606477853}{\strut flu} \colorbox{red!57.7558654}{\strut jab}\rule[-2ex]{0pt}{0pt} & R & BM25:\,3 \,\,\,\,\,  BERT\textsubscript{BM25CAT}:\,1 \,\,\,\,\,  BERT\textsubscript{CAT}:\,104 \\ 
    \arrayrulecolor[rgb]{0.753,0.753,0.753}\hline
    \rule{0pt}{4ex}\colorbox{red!0.0}{\strut [CLS]} \colorbox{red!11.892174062396489}{\strut what} \colorbox{blue!52.80559329547134}{\strut is} \colorbox{red!52.46300429290392}{\strut the} \colorbox{red!37.24586962843269}{\strut shingles} \colorbox{blue!70.38783808723808}{\strut jab} \colorbox{red!26.659793195189135}{\strut ?} \colorbox{red!0.0}{\strut [SEP]} \colorbox{blue!84.62420815046386}{\strut 11} \colorbox{blue!26.869425389989622}{\strut [SEP]} \colorbox{red!13.868294144786919}{\strut shingle} \colorbox{red!5.365354436476344}{\strut is} \colorbox{blue!4.5538655863916455}{\strut a} \colorbox{blue!5.075961707024493}{\strut corruption} \colorbox{blue!4.889461438595564}{\strut of} \colorbox{red!14.18672917029741}{\strut german} \colorbox{red!4.516496160120528}{\strut schindle} \colorbox{red!2.259669088055112}{\strut (} \colorbox{red!1.2822105198966574}{\strut schindel} \colorbox{blue!4.343224868439564}{\strut )} \colorbox{blue!3.4522554831297145}{\strut meaning} \colorbox{blue!3.7109858269852176}{\strut a} \colorbox{blue!1.5018226995863158}{\strut roofing} \colorbox{red!17.581523724714245}{\strut slate} \colorbox{red!9.466915952042289}{\strut .} \colorbox{blue!3.403252531235644}{\strut shingles} \colorbox{red!8.9813013034643}{\strut historically} \colorbox{red!5.486418606415349}{\strut were} \colorbox{red!15.681479981135976}{\strut called} \colorbox{blue!7.086856070324986}{\strut tiles} \colorbox{red!0.07441521495496924}{\strut and} \colorbox{blue!6.212463747458286}{\strut shingle} \colorbox{blue!2.0758987992952433}{\strut was} \colorbox{red!1.269241403990978}{\strut a} \colorbox{red!6.138468006765669}{\strut term} \colorbox{red!4.340403995753038}{\strut applied} \colorbox{red!8.175234815896928}{\strut to} \colorbox{blue!2.4928864142844827}{\strut wood} \colorbox{blue!5.682588977889988}{\strut shingles} \colorbox{red!4.596412128855955}{\strut ,} \colorbox{red!2.795867516096458}{\strut as} \colorbox{red!5.147749399895877}{\strut is} \colorbox{blue!1.4907053731991546}{\strut still} \colorbox{blue!0.43301381928355825}{\strut mostly} \colorbox{red!4.657057026684348}{\strut the} \colorbox{blue!6.5759031439897155}{\strut case} \colorbox{red!0.05566258828436476}{\strut outside} \colorbox{blue!11.353683643254293}{\strut the} \colorbox{blue!1.2203254387255833}{\strut us} \colorbox{red!0.0}{\strut [SEP]}  & N & BM25:\,146 \,\,\,\,\,  BERT\textsubscript{BM25CAT}:\,69 \,\,\,\,\,  BERT\textsubscript{CAT}:\,1 \\
    \arrayrulecolor{black}\hline
    \end{tabular}
}
\label{fig:casestudy}
\end{figure}
\par
\textbf{Qualitative analysis.}
We show a qualitative analysis of one particular case in Figure \ref{fig:casestudy} to analyze more in-depth what the effect of BM25 injection is and why it works. In the top row, while BERT\textsubscript{CAT} mistakenly ranked the relevant passage at position $104$, BM25 ranked that passage at position 3 and BERT\textsubscript{BM25CAT} -- apparently helped by BM25 -- ranked that relevant passage at position 1. In the bottom row, BERT\textsubscript{CAT} mistakenly ranked the irrelevant passage at position 1 and informed by the low BM25 score, BERT\textsubscript{BM25CAT} ranked it much lower, at 69. 
In order to interpret the importance of the injected BM25 score in the input of CE\textsubscript{BM25CAT} and show its contributions to the matching score in comparison to other words in the query and passage, we use Integrated Gradient (IG) \cite{sundararajan2017axiomatic} which has been proven to be a stable and reliable interpretation method in many different applications including Information Retrieval \cite{zhan2020analysis,chen2021toward,zhan2021interpreting}.
\footnote{We refer readers to \cite{sundararajan2017axiomatic} for a detailed explanation. }
On both rows of Figure \ref{fig:casestudy}, we see that the BM25 score (`22' in the top row and `11' in the bottom row) is a highly attributed term in comparison to other terms. This shows that injecting the BM25 score assists  BERT\textsubscript{BM25CAT} to identify relevant or non-relevant passages better than BERT\textsubscript{CAT}. 

As a more general analysis, we randomly sampled $100$ queries from MSMARCO-DEV. For each query, we took the top-$1000$ passages retrieved by BM25, we fed all pairs of query and their corresponding retrieved passages ($100k$ pairs) into BERT\textsubscript{BM25CAT}, and computed the attribution scores over the input at the word-level. We ranked tokens based on their importance using the absolute value of their attribution score and found the mode of the rank of the BM25 token over all samples is $3$. This shows that BERT\textsubscript{BM25CAT} highly attributes the BM25 token for ranking.
\section{Conclusion and future work}\label{sec:conclusion}
In this paper we have proposed an efficient and effective way of combining BM25 and cross-encoder re-rankers: injecting the BM25 score as text in the input of the cross-encoder. 
We find that the resulting model, CE\textsubscript{BM25CAT}, achieves a statistically significant improvement for all evaluated cross-encoders. Additionally, we find that our injection approach is much more effective than linearly interpolating the initial ranker and re-ranker scores. In addition, we show that CE\textsubscript{BM25CAT} performs significantly better in an exact matching setting than both BM25 and CE\textsubscript{CAT} individually. This suggests that injecting the BM25 score into the input could modify the current paradigm for training cross-encoder re-rankers.
\par
While it is crystal clear that our focus is not on chasing the state-of-the-art, we believe that as future work, our method could be applied into any cross-encoder in the current multi-stage ranking pipelines which are state-of-the-art for the MSMARCO Passage benchmark \cite{han2020learning}. Moreover, previous studies show that combining BM25 and BERT re-rankers on \textit{Robust04} \cite{allan2004overview} leads to improvement \cite{birch}. It is interesting to study the effect of injecting BM25 for this task because  documents often have to be truncated to fit the maximum model input length \cite{boytsov2022understanding}; injecting the BM25 score might give information to the cross-encoder re-ranker about the lexical relevance of the whole text of the document. Another interesting direction is to study how Dense Retrievers can benefit from injecting lexical ranker scores. Moreover, injecting scores of several lexical rankers and adding more traditional Learning-to-Rank features could be also interesting.

\section*{ACKNOWLEDGMENTS}
This work was supported by the EU Horizon 2020 ITN/ETN on Domain Specific Systems for Information Extraction and Retrieval (H2020-EU.1.3.1., ID: 860721).
\bibliographystyle{splncs04}
\bibliography{ref.bib}

\begin{thebibliography}{10}
\providecommand{\url}[1]{\texttt{#1}}
\providecommand{\urlprefix}{URL }
\providecommand{\doi}[1]{https://doi.org/#1}

\bibitem{abolghasemiICTIR}
Abolghasemi, A., Askari, A., Verberne, S.: On the interpolation of
  contextualized term-based ranking with bm25 for query-by-example retrieval.
  In: Proceedings of the 2022 ACM SIGIR International Conference on Theory of
  Information Retrieval. p. 161–170. ICTIR '22, Association for Computing
  Machinery, New York, NY, USA (2022). \doi{10.1145/3539813.3545133},
  \url{https://doi.org/10.1145/3539813.3545133}

\bibitem{abolghasemi2022improving}
Abolghasemi, A., Verberne, S., Azzopardi, L.: Improving bert-based
  query-by-document retrieval with multi-task optimization. In: European
  Conference on Information Retrieval. pp. 3--12. Springer (2022)

\bibitem{birch}
Akkalyoncu~Yilmaz, Z., Wang, S., Yang, W., Zhang, H., Lin, J.: Applying {BERT}
  to document retrieval with birch. In: Proceedings of the 2019 Conference on
  Empirical Methods in Natural Language Processing and the 9th International
  Joint Conference on Natural Language Processing (EMNLP-IJCNLP): System
  Demonstrations. pp. 19--24. Association for Computational Linguistics, Hong
  Kong, China (Nov 2019). \doi{10.18653/v1/D19-3004},
  \url{https://aclanthology.org/D19-3004}

\bibitem{al2022arabglossbert}
Al-Hajj, M., Jarrar, M.: Arabglossbert: Fine-tuning bert on context-gloss pairs
  for wsd. arXiv preprint arXiv:2205.09685  (2022)

\bibitem{allan2004overview}
Allan, J.: Overview of the trec 2004 robust retrieval track. In: Proceedings of
  TREC. vol.~13 (2004)

\bibitem{althammer2021dossier}
Althammer, S., Askari, A., Verberne, S., Hanbury, A.: {DoSSIER@ COLIEE 2021:
  Leveraging dense retrieval and summarization-based re-ranking for case law
  retrieval}. arXiv preprint arXiv:2108.03937  (2021)

\bibitem{anand2021serverless}
Anand, M., Zhang, J., Ding, S., Xin, J., Lin, J.: Serverless bm25 search and
  bert reranking. In: DESIRES. pp.~3--9 (2021)

\bibitem{askari2021combining}
Askari, A., Verberne, S.: Combining lexical and neural retrieval with
  longformer-based summarization for effective case law retrieva. In:
  Proceedings of the second international conference on design of experimental
  search \& information REtrieval systems. pp. 162--170. CEUR (2021)

\bibitem{AskariECIR22}
Askari, A., Verberne, S., Pasi, G.: Expert finding in legal community question
  answering. In: Hagen, M., Verberne, S., Macdonald, C., Seifert, C., Balog,
  K., N{\o}rv{\aa}g, K., Setty, V. (eds.) Advances in Information Retrieval.
  pp. 22--30. Springer International Publishing, Cham (2022)

\bibitem{bartell1994automatic}
Bartell, B.T., Cottrell, G.W., Belew, R.K.: Automatic combination of multiple
  ranked retrieval systems. In: SIGIR’94. pp. 173--181. Springer (1994)

\bibitem{berg2020empirical}
Berg-Kirkpatrick, T., Spokoyny, D.: An empirical investigation of
  contextualized number prediction. In: Proceedings of the 2020 Conference on
  Empirical Methods in Natural Language Processing (EMNLP). pp. 4754--4764
  (2020)

\bibitem{boualili2020markedbert}
Boualili, L., Moreno, J.G., Boughanem, M.: Markedbert: Integrating traditional
  ir cues in pre-trained language models for passage retrieval. In: Proceedings
  of the 43rd International ACM SIGIR Conference on Research and Development in
  Information Retrieval. pp. 1977--1980 (2020)

\bibitem{boualili2022highlighting}
Boualili, L., Moreno, J.G., Boughanem, M.: Highlighting exact matching via
  marking strategies for ad hoc document ranking with pretrained contextualized
  language models. Information Retrieval Journal pp. 1--47 (2022)

\bibitem{boytsov2022understanding}
Boytsov, L., Lin, T., Gao, F., Zhao, Y., Huang, J., Nyberg, E.: Understanding
  performance of long-document ranking models through comprehensive evaluation
  and leaderboarding. arXiv preprint arXiv:2207.01262  (2022)

\bibitem{chen2020numclaim}
Chen, C.C., Huang, H.H., Chen, H.H.: Numclaim: Investor's fine-grained claim
  detection. In: Proceedings of the 29th ACM International Conference on
  Information \& Knowledge Management. pp. 1973--1976 (2020)

\bibitem{chen2021toward}
Chen, L., Lan, Y., Pang, L., Guo, J., Cheng, X.: Toward the understanding of
  deep text matching models for information retrieval. arXiv preprint
  arXiv:2108.07081  (2021)

\bibitem{craswell2021overview}
Craswell, N., Mitra, B., Yilmaz, E., Campos, D.: Overview of the trec 2020 deep
  learning track. arXiv preprint arXiv:2102.07662  (2021)

\bibitem{craswell2021ms}
Craswell, N., Mitra, B., Yilmaz, E., Campos, D., Lin, J.: Ms marco:
  Benchmarking ranking models in the large-data regime. In: Proceedings of the
  44th International ACM SIGIR Conference on Research and Development in
  Information Retrieval. pp. 1566--1576 (2021)

\bibitem{craswell2020overview}
Craswell, N., Mitra, B., Yilmaz, E., Campos, D., Voorhees, E.M.: Overview of
  the trec 2019 deep learning track. arXiv preprint arXiv:2003.07820  (2020)

\bibitem{gao2021complement}
Gao, L., Dai, Z., Chen, T., Fan, Z., Durme, B.V., Callan, J.: Complement
  lexical retrieval model with semantic residual embeddings. In: European
  Conference on Information Retrieval. pp. 146--160. Springer (2021)

\bibitem{geva2020injecting}
Geva, M., Gupta, A., Berant, J.: Injecting numerical reasoning skills into
  language models. arXiv preprint arXiv:2004.04487  (2020)

\bibitem{ConceptNet2022}
Gretkowski, A., Wi{\'{s}}niewski, D., {\L}awrynowicz, A.: Should we afford
  affordances? injecting conceptnet knowledge into bert-based models to improve
  commonsense reasoning ability. In: Corcho, O., Hollink, L., Kutz, O.,
  Troquard, N., Ekaputra, F.J. (eds.) Knowledge Engineering and Knowledge
  Management. pp. 97--104. Springer International Publishing, Cham (2022)

\bibitem{gu-budhkar-2021-package}
Gu, K., Budhkar, A.: A package for learning on tabular and text data with
  transformers. In: Proceedings of the Third Workshop on Multimodal Artificial
  Intelligence. pp. 69--73. Association for Computational Linguistics, Mexico
  City, Mexico (Jun 2021). \doi{10.18653/v1/2021.maiworkshop-1.10},
  \url{https://www.aclweb.org/anthology/2021.maiworkshop-1.10}

\bibitem{han2020learning}
Han, S., Wang, X., Bendersky, M., Najork, M.: Learning-to-rank with bert in
  tf-ranking. arXiv preprint arXiv:2004.08476  (2020)

\bibitem{hofstatter2020improving}
Hofst{\"a}tter, S., Althammer, S., Schr{\"o}der, M., Sertkan, M., Hanbury, A.:
  Improving efficient neural ranking models with cross-architecture knowledge
  distillation. arXiv preprint arXiv:2010.02666  (2020)

\bibitem{johnson2020probing}
Johnson, D., Mak, D., Barker, D., Loessberg-Zahl, L.: Probing for multilingual
  numerical understanding in transformer-based language models. arXiv preprint
  arXiv:2010.06666  (2020)

\bibitem{kamphuis2020bm25}
Kamphuis, C., Vries, A.P.d., Boytsov, L., Lin, J.: Which bm25 do you mean? a
  large-scale reproducibility study of scoring variants. In: European
  Conference on Information Retrieval. pp. 28--34. Springer (2020)

\bibitem{khattab2020colbert}
Khattab, O., Zaharia, M.: Colbert: Efficient and effective passage search via
  contextualized late interaction over bert. In: Proceedings of the 43rd
  International ACM SIGIR conference on research and development in Information
  Retrieval. pp. 39--48 (2020)

\bibitem{kingma2014adam}
Kingma, D.P., Ba, J.: Adam: A method for stochastic optimization. arXiv
  preprint arXiv:1412.6980  (2014)

\bibitem{li2022markbert}
Li, L., Dai, Y., Tang, D., Feng, Z., Zhou, C., Qiu, X., Xu, Z., Shi, S.:
  Markbert: Marking word boundaries improves chinese bert. arXiv preprint
  arXiv:2203.06378  (2022)

\bibitem{li2002learning}
Li, X., Roth, D.: Learning question classifiers. In: COLING 2002: The 19th
  International Conference on Computational Linguistics (2002)

\bibitem{Lin_etal_SIGIR2021_Pyserini}
Lin, J., Ma, X., Lin, S.C., Yang, J.H., Pradeep, R., Nogueira, R.: {Pyserini}:
  A {Python} toolkit for reproducible information retrieval research with
  sparse and dense representations. In: Proceedings of the 44th Annual
  International ACM SIGIR Conference on Research and Development in Information
  Retrieval (SIGIR 2021). pp. 2356--2362 (2021)

\bibitem{pyseriniTunedBm25}
Lin, J., Ma, X., Lin, S.C., Yang, J.H., Pradeep, R., Nogueira, R.: Pyserini:
  Bm25 baseline for ms marco document retrieval (August 2021),
  \url{https://github.com/castorini/pyserini/blob/master/docs/experiments-msmarco-doc.md}

\bibitem{lin2021pretrained}
Lin, J., Nogueira, R., Yates, A.: Pretrained transformers for text ranking:
  Bert and beyond. Synthesis Lectures on Human Language Technologies
  \textbf{14}(4),  1--325 (2021)

\bibitem{macavaney2020expansion}
MacAvaney, S., Nardini, F.M., Perego, R., Tonellotto, N., Goharian, N.,
  Frieder, O.: Expansion via prediction of importance with contextualization.
  In: Proceedings of the 43rd International ACM SIGIR conference on research
  and development in Information Retrieval. pp. 1573--1576 (2020)

\bibitem{macavaney2019cedr}
MacAvaney, S., Yates, A., Cohan, A., Goharian, N.: Cedr: Contextualized
  embeddings for document ranking. In: Proceedings of the 42nd international
  ACM SIGIR conference on research and development in information retrieval.
  pp. 1101--1104 (2019)

\bibitem{ltrsolr}
Michael, N., Diego, C., Joshua, P., LP, B.: Learning to rank (May 2022),
  \url{https://solr.apache.org/guide/solr/latest/query-guide/learning-to-rank.html\#feature-engineering}

\bibitem{muffo-etal-2022-evaluating}
Muffo, M., Cocco, A., Bertino, E.: Evaluating transformer language models on
  arithmetic operations using number decomposition. In: Proceedings of the
  Thirteenth Language Resources and Evaluation Conference. pp. 291--297.
  European Language Resources Association, Marseille, France (Jun 2022),
  \url{https://aclanthology.org/2022.lrec-1.30}

\bibitem{nguyen2016ms}
Nguyen, T., Rosenberg, M., Song, X., Gao, J., Tiwary, S., Majumder, R., Deng,
  L.: Ms marco: A human generated machine reading comprehension dataset. In:
  CoCo@ NIPs (2016)

\bibitem{nogueira2019passage}
Nogueira, R., Cho, K.: Passage re-ranking with bert. arXiv preprint
  arXiv:1901.04085  (2019)

\bibitem{nogueira2019document}
Nogueira, R., Yang, W., Lin, J., Cho, K.: Document expansion by query
  prediction. arXiv preprint arXiv:1904.08375  (2019)

\bibitem{paszke2017automatic}
Paszke, A., Gross, S., Chintala, S., Chanan, G., Yang, E., DeVito, Z., Lin, Z.,
  Desmaison, A., Antiga, L., Lerer, A.: Automatic differentiation in pytorch
  (2017)

\bibitem{rau2022different}
Rau, D., Kamps, J.: How different are pre-trained transformers for text
  ranking? In: European Conference on Information Retrieval. pp. 207--214.
  Springer (2022)

\bibitem{rau2022role}
Rau, D., Kamps, J.: The role of complex nlp in transformers for text ranking.
  In: Proceedings of the 2022 ACM SIGIR International Conference on Theory of
  Information Retrieval. pp. 153--160 (2022)

\bibitem{reimers-2019-sentence-bert}
Reimers, N., Gurevych, I.: Sentence-bert: Sentence embeddings using siamese
  bert-networks. In: Proceedings of the 2019 Conference on Empirical Methods in
  Natural Language Processing. Association for Computational Linguistics (11
  2019), \url{https://arxiv.org/abs/1908.10084}

\bibitem{robertson2009probabilistic}
Robertson, S., Zaragoza, H., et~al.: The probabilistic relevance framework:
  Bm25 and beyond. Foundations and Trends{\textregistered} in Information
  Retrieval  \textbf{3}(4),  333--389 (2009)

\bibitem{robertson1994some}
Robertson, S.E., Walker, S.: Some simple effective approximations to the
  2-poisson model for probabilistic weighted retrieval. In: SIGIR’94. pp.
  232--241. Springer (1994)

\bibitem{salton1983introduction}
Salton, G., McGill, M.J.: Introduction to modern information retrieval.
  mcgraw-hill (1983)

\bibitem{sanh2019distilbert}
Sanh, V., Debut, L., Chaumond, J., Wolf, T.: Distilbert, a distilled version of
  bert: smaller, faster, cheaper and lighter. arXiv preprint arXiv:1910.01108
  (2019)

\bibitem{saracevic26review}
SARACEVIC, T.: A review of an a framework for the thinking on the notion in
  information science. Journal of the American Society for Information Science
  \textbf{26}

\bibitem{sundararajan2017axiomatic}
Sundararajan, M., Taly, A., Yan, Q.: Axiomatic attribution for deep networks.
  In: International conference on machine learning. pp. 3319--3328. PMLR (2017)

\bibitem{thawani2021representing}
Thawani, A., Pujara, J., Szekely, P.A., Ilievski, F.: Representing numbers in
  nlp: a survey and a vision. arXiv preprint arXiv:2103.13136  (2021)

\bibitem{vaswani2017attention}
Vaswani, A., Shazeer, N., Parmar, N., Uszkoreit, J., Jones, L., Gomez, A.N.,
  Kaiser, {\L}., Polosukhin, I.: Attention is all you need. In: Advances in
  neural information processing systems. pp. 5998--6008 (2017)

\bibitem{wallace2019nlp}
Wallace, E., Wang, Y., Li, S., Singh, S., Gardner, M.: Do nlp models know
  numbers? probing numeracy in embeddings. arXiv preprint arXiv:1909.07940
  (2019)

\bibitem{ZucconInterpolation}
Wang, S., Zhuang, S., Zuccon, G.: Bert-based dense retrievers require
  interpolation with bm25 for effective passage retrieval. In: Proceedings of
  the 2021 ACM SIGIR International Conference on Theory of Information
  Retrieval. p. 317–324. ICTIR '21, Association for Computing Machinery, New
  York, NY, USA (2021). \doi{10.1145/3471158.3472233},
  \url{https://doi.org/10.1145/3471158.3472233}

\bibitem{wang2020minilm}
Wang, W., Wei, F., Dong, L., Bao, H., Yang, N., Zhou, M.: Minilm: Deep
  self-attention distillation for task-agnostic compression of pre-trained
  transformers. Advances in Neural Information Processing Systems  \textbf{33},
   5776--5788 (2020)

\bibitem{wolf2019huggingface}
Wolf, T., Debut, L., Sanh, V., Chaumond, J., Delangue, C., Moi, A., Cistac, P.,
  Rault, T., Louf, R., Funtowicz, M., et~al.: Huggingface's transformers:
  State-of-the-art natural language processing. arXiv preprint arXiv:1910.03771
   (2019)

\bibitem{wu2009applying}
Wu, S.: Applying statistical principles to data fusion in information
  retrieval. Expert Systems with Applications  \textbf{36}(2),  2997--3006
  (2009)

\bibitem{Yan2019IDSTAT}
Yan, M., Li, C., Wu, C., Xia, J., Wang, W.: Idst at trec 2019 deep learning
  track: Deep cascade ranking with generation-based document expansion and
  pre-trained language modeling. In: TREC (2019)

\bibitem{yilmaz2019cross}
Yilmaz, Z.A., Yang, W., Zhang, H., Lin, J.: Cross-domain modeling of
  sentence-level evidence for document retrieval. In: Proceedings of the 2019
  conference on empirical methods in natural language processing and the 9th
  international joint conference on natural language processing (EMNLP-IJCNLP).
  pp. 3490--3496 (2019)

\bibitem{zhan2021interpreting}
Zhan, J., Mao, J., Liu, Y., Guo, J., Zhang, M., Ma, S.: Interpreting dense
  retrieval as mixture of topics. arXiv preprint arXiv:2111.13957  (2021)

\bibitem{zhan2020analysis}
Zhan, J., Mao, J., Liu, Y., Zhang, M., Ma, S.: An analysis of bert in document
  ranking. In: Proceedings of the 43rd International ACM SIGIR Conference on
  Research and Development in Information Retrieval. pp. 1941--1944 (2020)

\bibitem{zhang2020language}
Zhang, X., Ramachandran, D., Tenney, I., Elazar, Y., Roth, D.: Do language
  embeddings capture scales? arXiv preprint arXiv:2010.05345  (2020)

\bibitem{combiningScoresJimmyLin2021}
Zhang, X., Yates, A., Lin, J.: Comparing score aggregation approaches for
  document retrieval with pretrained transformers. In: Hiemstra, D., Moens,
  M.F., Mothe, J., Perego, R., Potthast, M., Sebastiani, F. (eds.) Advances in
  Information Retrieval. pp. 150--163. Springer International Publishing, Cham
  (2021)

\bibitem{zhang2021learning}
Zhang, Y., Hu, C., Liu, Y., Fang, H., Lin, J.: Learning to rank in the age of
  muppets: Effectiveness--efficiency tradeoffs in multi-stage ranking. In:
  Proceedings of the Second Workshop on Simple and Efficient Natural Language
  Processing. pp. 64--73 (2021)

\bibitem{zhang2018generalized}
Zhang, Z., Sabuncu, M.: Generalized cross entropy loss for training deep neural
  networks with noisy labels. Advances in neural information processing systems
   \textbf{31} (2018)

\bibitem{zhuang2021deep}
Zhuang, S., Li, H., Zuccon, G.: Deep query likelihood model for information
  retrieval. In: European Conference on Information Retrieval. pp. 463--470.
  Springer (2021)

\bibitem{zhuang2021tilde}
Zhuang, S., Zuccon, G.: Tilde: Term independent likelihood model for passage
  re-ranking. In: Proceedings of the 44th International ACM SIGIR Conference on
  Research and Development in Information Retrieval. pp. 1483--1492 (2021)

\end{thebibliography}
\end{document}